# Enhanced Photocurrent Response in Epitaxial 0.5PZT–0.5PFN Multiferroic Thin Films

Luc´ıa Imhoff[1,2], Miguel A. Rengifo[3,4,1*], Jos´e M. Caicedo Roque[2], Jessica Padilla-Pantoja[6], Jos´e Santiso[6], Marcelo G. Stachiotti[1], Myriam H. Aguirre[3,4,5]

[1] *IFIR - Instituto de F´ısica Rosario,*

*CONICET - Universidad Nacional de Rosario,*

*27 de Febrero 210 Bis, Rosario, Argentina*

[2] *FZU - Institute of Physics of the Czech Academy of Sciences, 182 00, Prague, Czechia*

[3] *Dept. de F´ısica de la Materia Condensada,*

*Universidad de Zaragoza, Pedro Cerbuna, 12, 50009, Zaragoza, Spain*

[4] *INMA - Instituto de Nanociencia y Materiales de Arag´on,*

*CSIC - Universidad de Zaragoza, Mariano Esquillor s/n, 50018, Zaragoza, Spain*

---

[1] *Laboratorio de Microscop´ıas Avanzadas, Universidad de Zaragoza,*

*Mariano Esquillor s/n, 50018, Zaragoza, Spain*

[2] *Catalan Institute of Nanoscience and Nanotechnology, ICN2,*

*CSIC and The Barcelona Institute of Science and Technology (BIST),*

*Campus UAB, Bellaterra, Barcelona, Spain*

# Abstract

The exploration of novel multiferroic materials with strong coupling between ferroelectric polarization and photovoltaic effects is crucial for next-generation optoelectronic devices. In this study, we characterized highly oriented $0.5Pb(Zr_{0.52}Ti_{0.48})O_3$–$0.5Pb(Fe_{0.5}Nb_{0.5})O_3$ multiferroic thin films grown by pulsed laser deposition on $SrTiO_3$ (001) substrates with a $SrRuO_3$ bottom electrode. The films exhibited excellent crystalline quality, with a single perovskite phase and (001) orientation. They displayed good ferroelectric properties (remanent polarization ≈ 17 $\mu C/cm^2$, PUND; coercive field ≈ 150 kV/cm), alongside weak ferromagnetic behavior at room temperature (remanence 1.30 $emu/cm^3$; coercive field 90 Oe). Photovoltaic measurements demonstrated a robust, polarization-dependent photoresponse under 403 nm monochromatic laser illumination, achieving $J_{sc}$ values between ±20 $\mu A/cm^2$. This compelling observation confirms the intrinsic coupling between ferroelectric polarization and photovoltaic effects, highlighting the considerable promise of these single-phase multiferroic thin films for advanced photovoltaic and optoelectronic memory applications.

* marengifom@unizar.es.

## I. INTRODUCTION

Multiferroic materials have garnered significant attention over the past few decades and remain a focal point of current research. They can be basically defined as materials that simultaneously exhibit multiple ferroic properties within the same temperature range, including ferroelectricity, ferromagnetism, and ferroelasticity. The unique combination of these properties makes multiferroics attractive for a wide range of applications, from advanced memory devices to novel sensors and energy storage [1–3]. Furthermore, the ability to couple magnetic and electric orders would enable the control of magnetic properties through electric fields and vice versa, opening new opportunities for low-power spintronics and energy-efficient data storage [3]. Beyond the mentioned applications, multiferroics present a significant potential in the field of ferroelectric photovoltaics, as the presence of d-states in magnetic ions tends to reduce the band gap energy, enhancing the absorption of photons at lower energies [4]. For instance, $BiFeO_3$ (BFO) thin films exhibit a narrow band gap (˜2eV), which enables efficient photon harvesting and makes them promising candidates
for photovoltaic absorbers [5].

While single-phase multiferroics with robust polarization and magnetism remain rare, BFO has been dominating research efforts in the field. On the other hand, given its exceptional ferroelectric behavior and widespread market use, it is also worthwhile to explore lead zirconate titanate (PZT) - based materials as potential multiferroics. In particular, $Fe^{3+}/Nb^{5+}$ ($Ta^{5+}$) co-doping of PZT through the fabrication of (1-x)PZT-xPFN (PFT)
solid solutions has proven to be a successful strategy for obtaining single-phase multiferroic materials in both bulk [6, 7] and thin film [8–11] forms. In bulk ceramics, compositions with x = 0.2-0.4 exhibit saturated ferroelectric loops (remanent polarization $P_r \approx$ 20–30 μC/cm$^2$) and enhanced magnetization at x = 0.2 and x = 0.3 [6], with clear evidence of
magnetic-field-induced domain switching [7]. All these studies demonstrate that the PZTPFN system represents a viable alternative to BFO as a room-temperature multiferroic. Thin-film studies are scarcer but reveal relaxor-like ferroelectricity ($P_r \approx$ 13–20 μC/cm$^2$) and weak ferromagnetism (remanent magnetization $M_r \approx$ 0.2–1.6 emu/cm$^3$) in highly oriented $Pb(Zr_{0.53}Ti_{0.47})_{0.60} - (Fe_{0.5}Ta_{0.5})_{0.40}O_3$/LSCO//MgO(001) samples [9–11]. The weak ferromagnetic behavior of the films is attributed to super-exchange interactions through $Fe^{3+}$-O-$Fe^{3+}$, along with tensile strain that may induce a slight distortion in the spins. It is worth noting that Fe/Nb co-doping presents an advantage over Fe/Ta: in bulk samples, Fe/Nb-doped PZT exhibited higher Pr and Curie temperature values [6, 12, 13]. However, there are no previous reports of studies conducted on epitaxial Fe/Nb co-doped PZT films. In previous works, our group synthesized polycrystalline (1-x)PZT-xPFN (x = 0–0.5) thin films using a modified sol-gel route [8]. The films showed improved magnetization loops for x = 0.4 and x = 0.5 ($M_r \approx$ 2 emu/cm$^3$, $H_c \approx$ 1500 Oe for x = 0.5), along with well-saturated ferroelectric loops for all studied concentrations ($P_r \approx$ 10 μC/cm² and $E_c \approx$ 80 kV/cm for x = 0.5). Additionally, we demonstrated a significant enhancement of the optical absorption in the visible light region, by reducing the direct band gap energy from 3.6 eV in PZT to 3.0 eV in 0.5PZT-0.5PFN, making this solid solution a promising candidate for multiferroic-photovoltaic applications [14].

In this work, we report the first comprehensive study of epitaxial $0.5Pb(Zr_{0.52}Ti_{0.48})O_3$–$0.5Pb(Fe_{0.5}Nb_{0.5})O_3$ ($PZTFN_{0.5}$) thin films grown by pulsed laser deposition (PLD) on $SrTiO_3$(001) substrates with $SrRuO_3$ (SRO) as a bottom electrode. Through detailed structural, ferroelectric, magnetic, and photovoltaic characterization, we demonstrate the material's multifunctional capabilities for ferroelectric photovoltaics (FPV). Notably, the short-circuit photocurrent ($J_{sc}$) under 403 nm illumination exhibits strong modulation with both poling direction and magnitude of the applied electric field, revealing direct coupling between ferroelectric polarization and photovoltaic response.

## II. EXPERIMENTAL

A stoichiometric ceramic target of $0.5Pb(Zr_{0.52}Ti_{0.48})O_3$–$0.5Pb(Fe_{0.5}Nb_{0.5})O_3$ was synthesized via the solid-state reaction method. Precursors PbO (99.9%), $ZrO_2$ (99.9%), $TiO_2$ (99%), $Fe_2O_3$ (99%) and $Nb_2O_5$ (99.99%) were weighed to achieve a nominal composition where Zr, Ti, Fe and Nb each constitute 5 at.% of the B-site cations in the perovskite structure. The powders were mixed with acetone and hand-milled in an agate mortar for 30 min. A 10 wt.% PbO excess was added to compensate for losses due to volatilization. The mixture was calcined at 850 °C for 6 h, then homogenized with Polaroid® binder and re-milled for 30 minutes. The powder was uniaxially pressed into pellets at an 8-ton load and sintered at 950 °C for 2 h in a sealed crucible with a $PZTFN_{0.5}$ powder bed to minimize Pb loss. A low heating ramp (1 °C/min) was applied up to 600 °C to prevent cracking. Epitaxial thin films were fabricated by pulsed laser deposition (PLD) using this target. For comparison, pure PZT thin films were also grown under identical conditions using a commercial $Pb(Zr_{0.52}Ti_{0.48})O_3$ target (Kurt J. Lesker).

The films were grown onto STO (001)-oriented (Crystal, GmbH.) substrates. The ablation was performed using a KrF excimer laser (Lambda Physik COMPex 201, $\lambda$ = 248 nm wavelength). The SRO conductive layer was deposited at 635 °C, and both PZT and $PZTFN_{0.5}$ films were grown at 550 ∘C. This was found to be the optimum temperature for stabilizing the perovskite phase, as traces of secondary pyrochlore phases appeared in $PZTFN_{0.5}$ samples treated at higher temperatures (600 ∘C). Deposition conditions were: 100 mTorr $O_2$ pressure, laser fluence of 1.5–2 J/cm$^2$, and target-to-substrate distance of 55 mm. The cooling down step was at 10 ∘C/min, keeping a 10 Torr $O_2$ pressure. The final thickness of the samples was around 80 nm (growth rate ≈ 0.04 nm/pulse). Circular Au top electrodes of 200 and 100 $\mu$m diameter (50 nm thick) were deposited by thermal evaporation through a shadow mask to fabricate micro-capacitors for electrical characterization of the films.

The crystallinity of the films was evaluated by x-ray diffraction by measuring $\theta$–2$\theta$ scans and reciprocal space mapping (RSM). A diffractometer with a 4-angle goniometer (MRD X'Pert Pro, Malvern-Panalytical) equipped with a Cu-K$\alpha$ ($\lambda$ = 1.540 ˚A) tube was used. The detector employed was a PIXcel with a 256 × 256 pixel array. The microstructural details of the fabricated heterostructures were analyzed by Scanning Transmission Electron Microscopy (STEM) operated in High-Angle Annular Dark Field (HAADF) mode. A FEI TitanG2 80–300 microscope, coupled to a Fischione (HAADF-STEM) detector, was operated at 300 kV at room temperature with a probe-corrected beam (CEOS Co.). This technique was used to determine the thickness of the films, as well as to examine their microstructure and chemical composition. Strain maps were calculated from STEM-HAADF images using General Phase Analysis (GPA) software.

Electrical measurements were performed at room temperature with a shielded probe station equipped with a Faraday cage to minimize electromagnetic interference and ensure completely dark conditions. DC electrical characterization was carried out using a Keithley 2636B SourceMeter unit, while capacitance–voltage and dielectric measurements were conducted with an Agilent 4294A precision impedance analyzer. Ferroelectric polarization–electric field ($P$–$E$) hysteresis loops were characterized using a Sawyer–Tower circuit driven by a Keysight 33210A function generator (1 kHz triangular waveform) and monitored with a Rigol DHO924S high-resolution oscilloscope. Data were acquired and processed to determine the remanent polarization ($P_r$) and coercive field ($E_c$).

The photovoltaic response was characterized under both dark and illuminated conditions using a 403 ± 8 nm ($h\nu$ = 3.08 eV) laser source with a constant calibrated optical fluence of 100 ± 5 mW/cm$^2$. Magnetic measurements were carried out using a MPMS XL (Magnetic Properties Measurement System, Quantum Design). The surface roughness and ferroelectric response were further investigated using Piezo Response Force Microscopy (PFM) on a Veeco Multimode 8 Atomic Force Microscopy (AFM) system.

## III. RESULTS

### A. Structural properties and nanomorphology

**Figure 1a** shows the room-temperature x-ray diffraction patterns of the analyzed thin films. The pattern obtained for the PZT sample is also shown as a reference. The measurements were made with $2\theta$ varying between 10$^\circ$ and 90$^\circ$. The films present a single-phase perovskite crystalline structure epitaxially grown along the (001) direction, driven by the orientation of the STO substrate. The shoulders appearing at lower angles of the STO peaks correspond to the SRO layer that grows coherently strained to the substrate. No signs of different crystalline orientations nor secondary phases are observed. The appearance of layer fringes below the (001) peak of $PZTFN_{0.5}$ evidences the excellent crystalline quality of the SRO layer, along with a sharp interface. The SRO layer thickness determined from the period of the oscillations is approximately 20 nm for both samples.

To further characterize the crystalline quality and strain behavior of the layers, reciprocal space maps (RSM) were measured using the $(\bar{1}03)$ reflection direction. The fully textured out-of-plane growth of the pseudo-cubic films (the epitaxial relation between the film and the substrate) could also be confirmed by this technique. As shown in **Figure 1b**, the SRO conductive bottom layer coherently grows on the STO substrate with in-plane parameter $a$ = 3.905 Å and out-of-plane parameter $c$ = 3.945 Å. The $PZTFN_{0.5}$ film layer appears to grow completely relaxed with lattice parameters $a$ = 4.022 Å, $c$ = 4.096 Å, and $c/a$ = 1.018. This tetragonality closely aligns with that observed in bulk ceramics, which exhibit a $c/a$ ratio of 1.016 [8]. A simple calculation of the lattice mismatch between the substrate and the deposited material can be done using the following expression in-line: $\mu = (a_s - a_{\mathrm{film}})/a_s$, where $a_s$ refers to the in-plane lattice parameter of the STO substrate and $a_{\mathrm{film}}$ refers to the in-plane lattice parameter of the unstrained material. In our case, the mismatch between the substrate (STO) and the film gives a value of around 3%, corresponding to compressive strain.

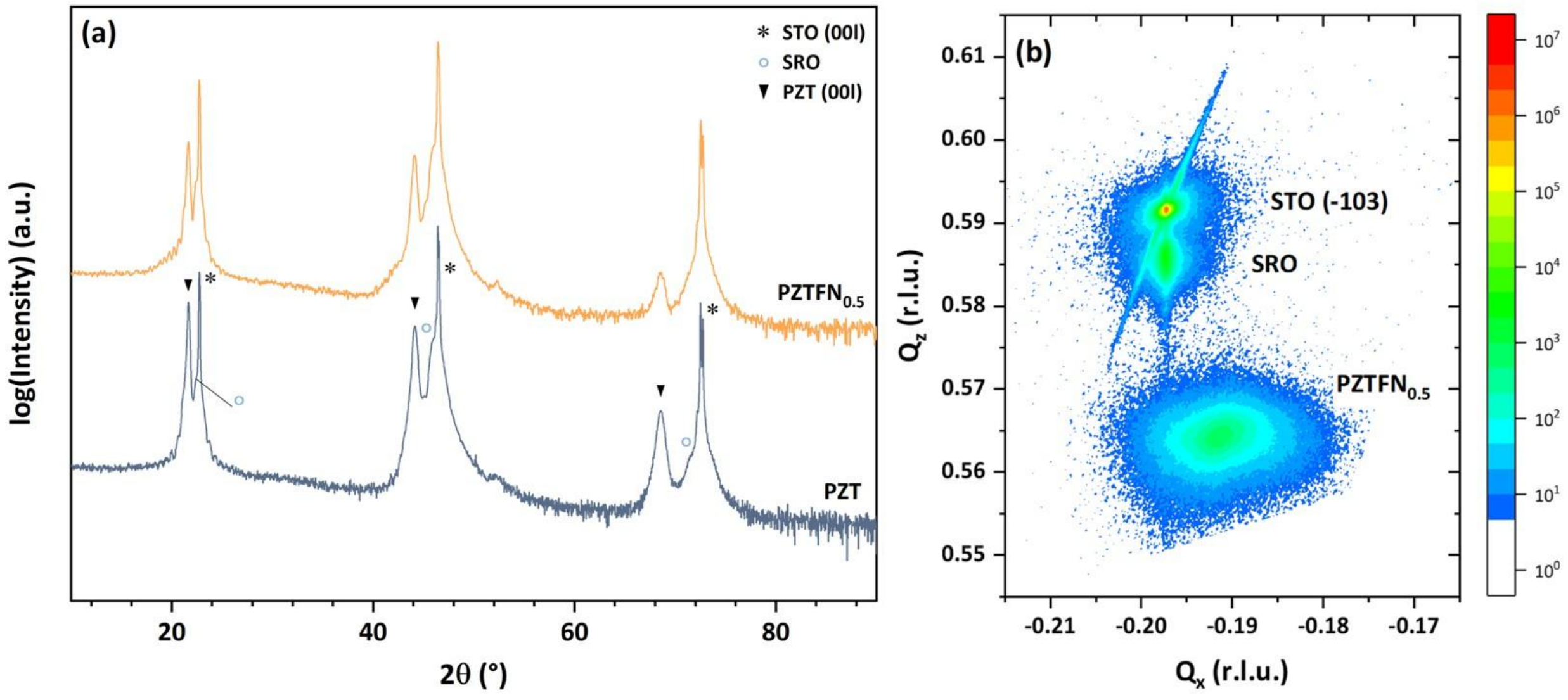

**Figure 1**. **(a)** Room-temperature XRD patterns of the STO//SRO/$PZTFN_{0.5}$ and STO//SRO/PZT thin films. **(b)** Reciprocal space map (RSM) measured in the $(\bar{1}03)$ reflection direction for the $PZTFN_{0.5}$ sample.

**Figure 2** presents the transmission electron microscopy (TEM) characterization of the $PZTFN_{0.5}$ film. Cross-sectional HAADF-STEM images were obtained. In the lowmagnification images (**Fig. 2(a, b)**), the complete profile of the film is visible, with the different layers indexed and their respective thicknesses indicated. The excellent crystalline quality of the sample is also confirmed by this technique.

Using GPA software, we analyzed the strain field map of $PZTFN_{0.5}$ deposited on the SRO layer (**Fig. 2(b, c)**). Strain maps for the $\varepsilon_{xx}$ component, which corresponds to variations in the in-plane lattice parameter, were generated using the STO substrate as a reference. The film grows in a fully relaxed state beyond a clamped region only a few nanometers thick (**Fig. 2c**), where a line of dislocations (indicated by dashed circles) is observed. This region is located right at the interface between the SRO layer and the film. Up to that point, no significant deformation is detected, indicating no notable change in the in-plane lattice parameter.

A high-resolution image of the SRO–$PZTFN_{0.5}$ interface is shown in **Fig. 2d**, where dislocations within the clamped region are clearly visible, as highlighted by a dashed circle. The relaxed state observed in the GPA strain map is consistent with the x-ray reciprocal space map (RSM) measurements previously discussed (**Fig. 1b**). Additional compositional information obtained from STEM-EDS line profile measurements across the heterostructure is provided in the Supplementary Information (SI) (**Figure S1**). Finally, in **Figure S2** of the SI, we present TEM and STEM-HAADF images for the STO//SRO/PZT thin film, the sample used as a reference throughout this work.

### B. Ferroelectric and magnetic properties

The ferroelectric properties were characterized using Capacitance–Voltage (C–V) and Polarization–Electric Field (P–E) hysteresis loop measurements in a conventional bottom–top electrode configuration. In this arrangement, the bottom $SrRuO_3$ (SRO) layer acted as the bottom electrode, while the top Au electrode (positively biased) served as the top electrode, forming a metal–ferroelectric–metal (MFM) capacitor structure with the $PZTFN_{0.5}$ film as the dielectric layer (see inset in **Figure 3a**). We defined positive polarization (P-Up) as the state where the polarization vector points toward the top Au electrode (positively biased), and negative polarization (P-Down) as the state where the vector points toward the bottom SRO electrode (negatively biased).

For C–V measurements, an AC signal amplitude of 0.2 V at 10 kHz was applied. For P–E hysteresis loops, a triangular waveform with an alternating electric field was used at 1 kHz and an amplitude of 6 $V_{pp}$. For both PZT and $PZTFN_{0.5}$, the C–V and P–E loops are shown in **Figure 3a** and **3b**, respectively. The C–V measurements exhibit the classical butterflyshaped curve, while the P–E loops display the characteristic hysteresis behavior expected for ferroelectric materials. The C–V curves for both samples were normalized to their respective maximum capacitance values: 690 pF for PZT and 960 pF for $PZTFN_{0.5}$. Both C–V and P–E loops exhibit a positive imprinted polarization state, indicating a significant internal built-in field ($E_{in}$) directed toward the top electrode. The estimated imprint field for the $PZTFN_{0.5}$ sample from P–E loops is approximately 40 kV/cm (**Fig. 3b**). In this regard, the loops can be considered asymmetrical, particularly for the Fe/Nb-substituted sample, which also exhibits weaker saturation at positive fields. A similar, though less pronounced, issue is observed in the PZT loop. The imprint phenomenon may arise from several factors, including asymmetric electrode–ferroelectric interfaces, stress from lattice mismatch or substrate clamping, domain pinning due to defect–dipole alignments, and/or a graded surface layer [15, 16]. As discussed earlier, TEM images revealed a clamped region at the interface between the bottom SRO electrode and the ferroelectric thin film (**Fig. 2**).

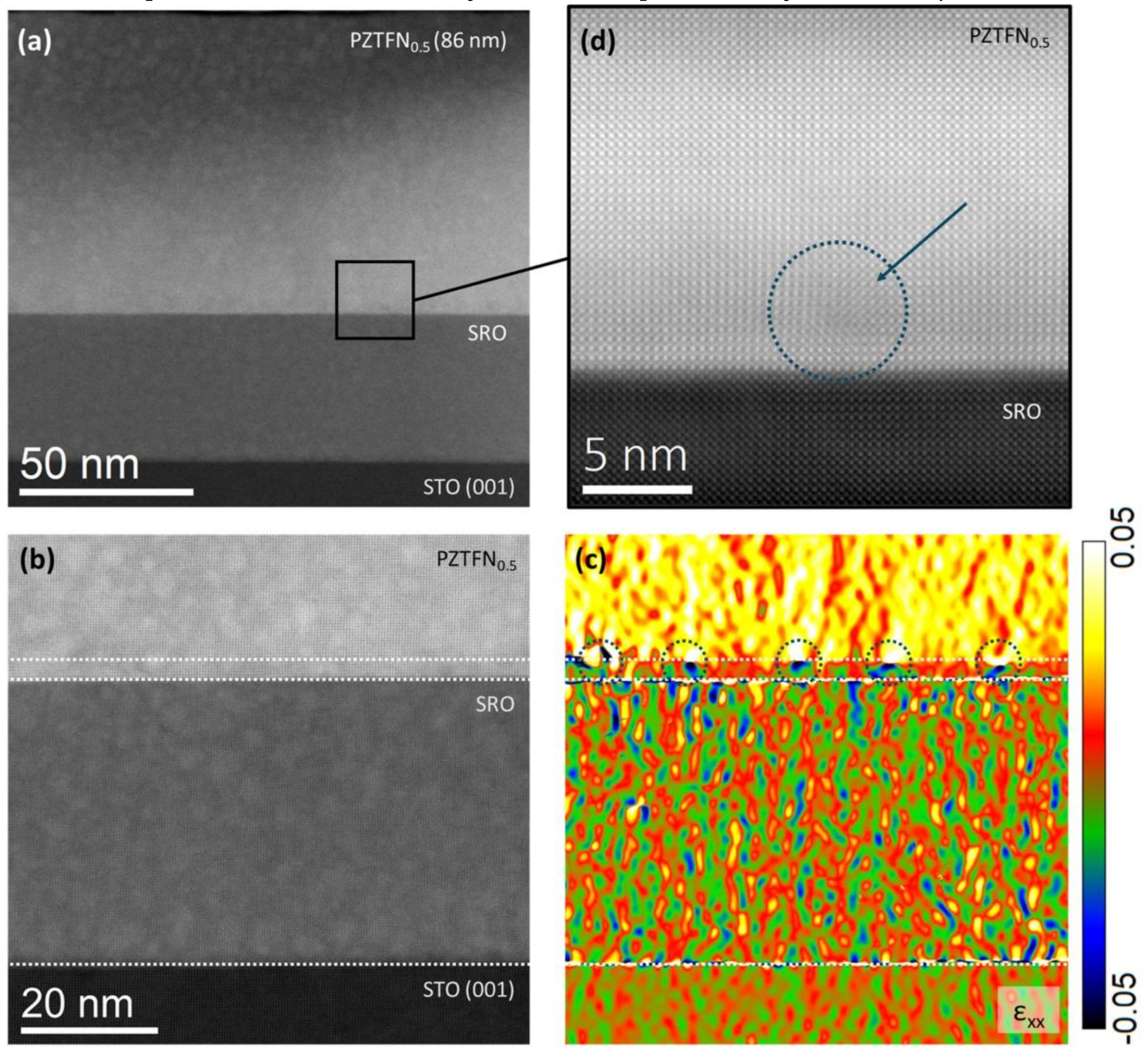

**Figure 2**. **(a)** Cross-sectional STEM-HAADF image of the $PZTFN_{0.5}$ sample, showing the SRO conductive layer and STO (001) substrate. **(b, c)** Cross-sectional image and strain ($\varepsilon_{xx}$) component map for the $PZTFN_{0.5}$ film grown on an STO (001) substrate with an SRO conductive layer. **(d)** High resolution image of the interface between the film and the SRO layer. A dislocation is highlighted by the dashed circle.

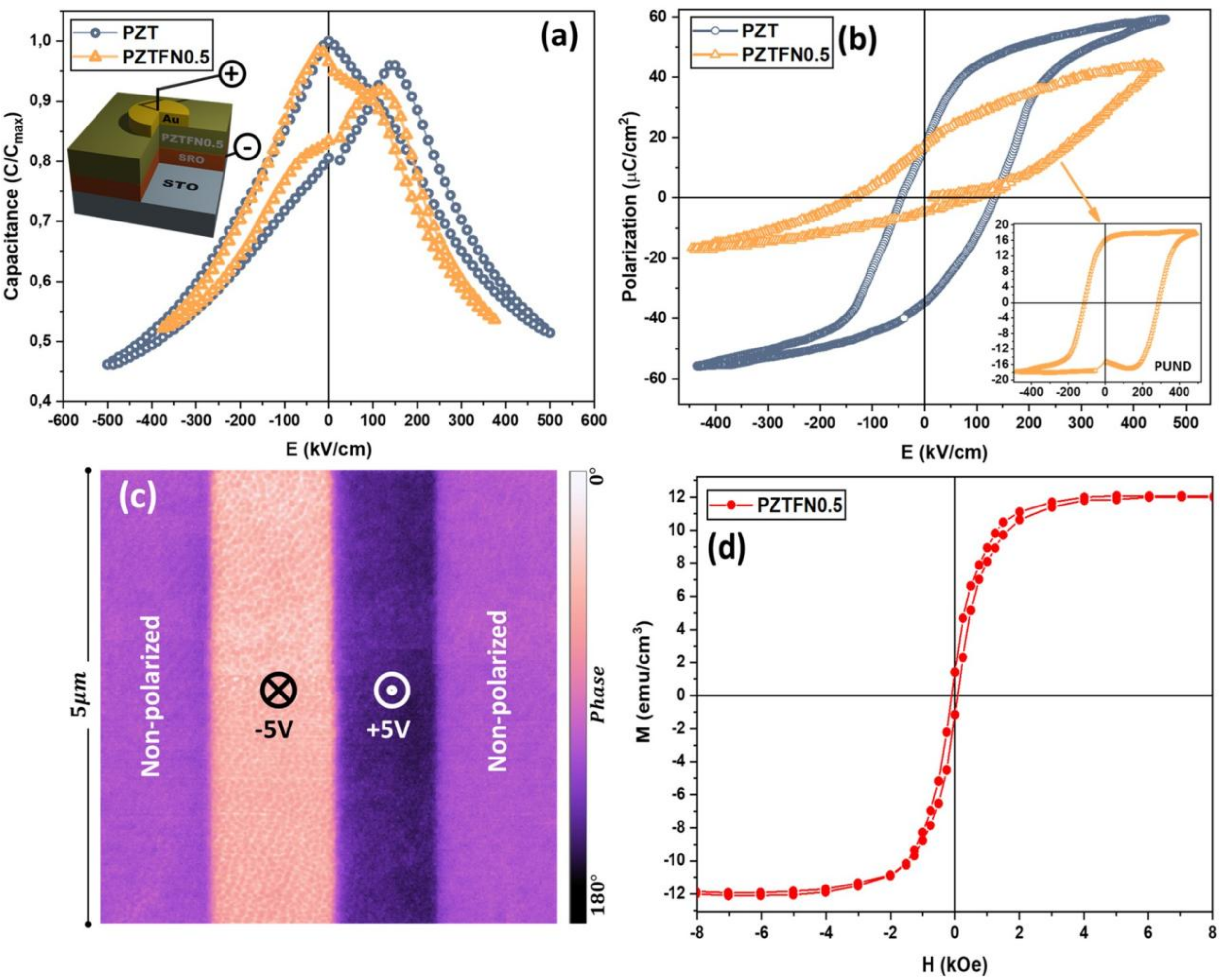

**Figure 3**. **(a)** Normalized Capacitance–Voltage (C–V) and **(b)** Polarization–Electric Field (P–E) loops for STO//SRO/PZT/Au and STO//SRO/$PZTFN_{0.5}$/Au thin films. **(c)** PFM phase image for the STO//SRO/$PZTFN_{0.5}$ system. **(d)** M–H loop measured at room temperature for the STO//$PZTFN_{0.5}$ thin film.

This could be a possible explanation for the observed imprint, along with the asymmetry caused by different electrode materials (different work functions).

It is known that if B-site cations $Fe^{3+}$ and $Nb^{5+}$ are not well compensated, defects can form (either positive or negative charges) [16]. In general, the presence of defects hardens domain wall motion, which could explain the higher coercive field observed in the substituted sample compared to PZT [17]. On the other hand, the saturation polarization reached in PZT is about 30% higher than in the $PZTFN_{0.5}$ sample. These results indicate that Fe/Nb substitution slightly reduces the ferroelectric behavior in epitaxial samples, in contrast to polycrystalline samples, where ferroelectric parameters remained mostly unchanged with increasing substitution ions [8].

The ferroelectric properties of $PZTFN_{0.5}$ were further investigated using the Positive-UpNegative-Down (PUND) technique, which effectively isolates the switching polarization from non-ferroelectric contributions such as leakage currents and dielectric capacitance [18, 19]. The measurement protocol employed consecutive unipolar triangular pulses at 1 kHz with the following sequence: two positive pulses (P: Positive, U: Up) to establish saturated polarization in the "up" state, followed by two negative pulses (N: Negative, D: Down) to switch the polarization to the "down" state. The true ferroelectric polarization was extracted by calculating the difference between switched (P–U) and non-switched (N–D) responses, effectively eliminating parasitic conduction effects. This differential analysis yields a characteristic square hysteresis loop (**Inset Fig. 3b**) that represents the intrinsic ferroelectric switching behavior, with the vertical displacement corresponding to the remanent polarization ($P_r$). The $P_r$ value obtained using this technique was 17 $\mu$C/cm$^2$. To evaluate the stability of the ferroelectric states, retention measurements were performed using the PUND method. **Fig. S5** shows that the Up and Down polarization states remain clearly distinguishable up to $10^4$ s. A gradual reduction of the remanent polarization is observed with time, which may be related to depolarization processes, interface defects, and slow charge screening effects. Since the photovoltaic response is strongly dependent on the polarization orientation, the retention of the ferroelectric states supports the stability of the polarization-dependent photoresponse.

The obtained polarization values for the $PZTFN_{0.5}$ film are comparable to those reported for similar compositions. For instance, $PZTFT_{40}$/LSCO//MgO(001) films exhibit a remanent polarization of 13 $\mu C/cm^2$ [9]. However, the coercive field is significantly lower in this case ($E_c$ = 40 kV/cm). Polycrystalline $PZTFN_{0.5}$ films deposited on Pt/Ti/$SiO_2$/Si substrates have reported $P_r$ and $E_c$ values of 10 $\mu C/cm^2$ and 80 kV/cm, respectively [8].

We further characterized the ferroelectric properties through piezoresponse force microscopy (PFM) measurements. **Figure 3c** reveals nearly complete ferroelectric domain alignment with ±5 V in the $PZTFN_{0.5}$ sample, exhibiting distinct phase contrast between P-Up and P-Down polarization states as well as non-polarized regions. The measured surface roughness was approximately 0.82 nm, confirming the high quality of the sample and
its suitability for nanoscale ferroelectric studies.

To investigate multiferroic features at room temperature, in-plane magnetization values were collected as a function of the applied magnetic field $H$. The measurements were performed on a film deposited directly on an STO substrate (STO//$PZTFN_{0.5}$), to avoid possible contributions from the magnetic response of the SRO layer [20]. The microstructural (XRD + TEM) characterization of the corresponding sample is shown in the SI, **Figures S3** and **S4**, respectively. The results are presented in **Figure 3d** as a magnetization–field (M– H) hysteresis loop. The magnetization and coercive field values measured for the $PZTFN_{0.5}$ epitaxial film ($M_r$ = 1.3 emu/$cm^3$, $H_c$ = 90 Oe) are quite similar to those reported for $PZTFT_{0.4}$ epitaxial films ($M_r$ = 1.58 emu/$cm^3$, $H_c$ = 108 Oe) [9]. Interestingly, a magnetic saturation of 12 emu/$cm^3$ is observed, which is higher than that for $PZTFT_{0.4}$ films (7 emu/$cm^3$). When comparing our PLD-deposited $PZTFN_{0.5}$ films with sol-gel polycrystalline films, the remanences are similar (1.3 emu/$cm^3$ and 1.8 emu/$cm^3$ [14]), but there is
a significant difference in coercive field values (90 Oe versus 1500 Oe).

### C. Photovoltaic response

The photovoltaic response was characterized through current density ($J$) measurements under dark and illuminated conditions using monochromatic laser excitation at $\lambda$ = 403 nm (3.07 eV, violet). This wavelength was selected to probe the material's response at 403 nm with a direct band gap of 3.0 eV and further validate our previous findings on $PZTFN_{0.5}$ polycrystalline thin films [14]. The laser beam was focused through the top electrode with a spot size larger than the electrode diameter, as shown in the inset in **Figure 4a**, with a calibrated irradiance for all measurements. The measurements were conducted at room temperature without any pre-poling treatment. The *J*–*E* response measured for applied voltages ranging from −3 V to 3 V (±375 kV/cm) is shown in **Figure 4a**. The devices underwent breakdown at applied voltages ≥ 4.0 V for these DC measurements. The asymmetric *J*–*E* characteristics further confirm the inherent structural asymmetry of the device
architecture, consistent with our previous observations.

A notable increase in conductivity (about one order of magnitude) occurs when the light is switched ON compared to the OFF state. This behavior indicates that the $PZTFN_{0.5}$ layer effectively absorbs light at 403 nm, resulting in the generation of free charge carriers. Under dark conditions, an important hysteretic behavior can be observed at negative voltage values. These results suggest enhanced modulation at negative bias voltages, likely originating from interface effects at the SRO/$PZTFN_{0.5}$ junction. Under illuminated conditions (light-ON), no hysteretic behavior was observed. This behavior contrasts sharply with our STO//SRO/PZT/Au reference system, which exhibits negligible optical absorption in this spectral range (see SI, **Figure S6**). This is consistent with the reported direct band gap of ~3.6 eV [14]] for PZT, which requires higher-energy photons for carrier generation. Based on these findings, we focus subsequent analysis exclusively on the SRO/$PZTFN_{0.5}$/Au heterostructure. A higher temporal resolution inset of the 403 nm transient photocurrent response in figure **Figure S6a** shows a fast initial excitation process occurring within a few milliseconds, followed by slower saturation and relaxation processes. These observations suggest that charge trapping and interface-related effects may influence the photocarrier
transport and recombination processes in the films.

**Figure 4b** displays current density–electric field (*J*–*E*) characteristics measured at low bias voltages (±0.3 V) to determine the short-circuit photocurrent density ($J_{sc}$ at $V$ = 0) and open-circuit voltage ($V_{oc}$ at $I$ = 0) under 403 nm laser illumination. These lowvoltage measurements ensure that the polarization state remains unaltered. Prior to each measurement, distinct polarization states were initialized by applying ±3 V for 2 minutes in dark conditions. The measurements reveal polarization-dependent photocurrent generation, with positive $J_{sc}$ for the P-Up state and negative $J_{sc}$ for the P-Down state.

This polarization-dependent behavior correlates with the relative permittivity ($\varepsilon'_r$) trends shown in **Figure 4c**. While $\varepsilon'_r$ remains stable at low frequencies (20 Hz–10 kHz), significant polarization-state-dependent variations emerge at higher frequencies (>10 kHz). This frequency dependence reflects the transition from interface-dominated effects (electrode contributions in the metal–insulator–metal structure at low frequencies) to bulkdominated behavior (grain boundary and ferroelectric domain responses at high frequencies) [21]. Our results therefore demonstrate dominant bulk ferroelectric effects in the measured response. This particular behavior for $\varepsilon_r$ was also reported in $(K_{0.5}Na_{0.5})NbO_3$–2 mol% $Ba(Ni_{0.5}Nb_{0.5})O_{3-\delta}$ samples [22].

To elucidate the relationship between photovoltaic response and ferroelectric polarization, we systematically investigated the dependence of $J_{sc}$ and $V_{oc}$ on poling voltage (**Figure 4d**). For each data point in this figure, the sample was pre-poled for 2 minutes at different voltage values to establish specific polarization states, following the methodology illustrated in **Figure 4b**. The $J_{sc}$ and $V_{oc}$ were subsequently measured immediately after each poling procedure to ensure consistent polarization conditions. The results demonstrate a clear correlation between polarization state and photovoltaic output: the P-Up state consistently yields positive $J_{sc}$ and negative $V_{oc}$ values, while the P-Down state produces the opposite behavior. This robust polarization-dependent photoresponse further confirms the intrinsic coupling between ferroelectric domains and photovoltaic effects in the material. These results clearly demonstrate the influence of the imprint electric field identified in our prior ferroelectric characterization, consistent with the asymmetric P–E behavior observed in **Figure 3a** and **3b**.

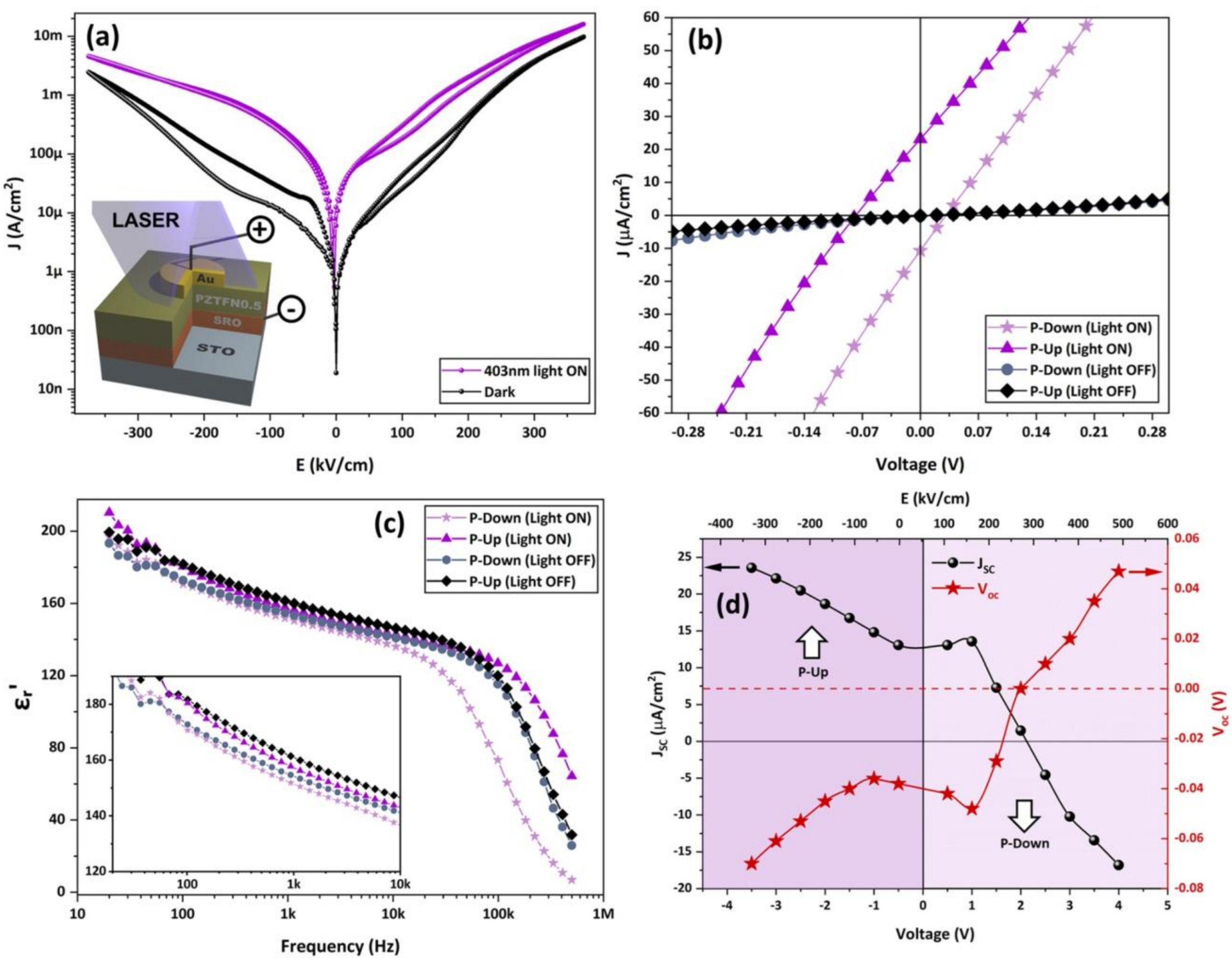

**Figure 4**. **(a)** $J$–$E$ response measured in the STO//SRO/PZTFN$_{0.5}$/Au sample under dark conditions and under illumination with 430 nm light. **(b)** $J$–$E$ response at low electric fields ($E$ < $E_c$). **(ac)** $\varepsilon'_r$ variation under different polarization and illumination conditions. **(d)** $J_{SC}$ and $V_{OC}$ values under different poling voltages.

We propose a simplified physical model to explain the observed behavior across all experimental results. The model assumes that PZTFN$_{0.5}$ behaves as an n-type semiconductor. The electron affinities involved are $e\phi_{SRO}$ = 4.6 eV [23] and $e\phi_{Au}$ = 5.1 eV [24], considering a band gap $E_g$ = 3.0 eV for PZTFN$_{0.5}$ [14]. These parameters were used to construct the equilibrium energy band diagram shown in **Figure 5a**, where the electronic structure of each material is first represented separately before analyzing the interface formation.

At equilibrium, the heterostructure comprising SRO/$PZTFN_{0.5}$/Au forms two back-toback Schottky junctions: D1 ($PZTFN_{0.5}$/Au) with a barrier height of 1.2 eV, and D2 (SRO/$PZTFN_{0.5}$) with a barrier height of 0.7 eV. This configuration arises from Fermi-
level alignment across the interfaces, creating the characteristic double-diode behavior, as shown in **Figure 5b**.

When a positive external bias ($V_{ext} > 0$) is applied, as shown in **Figure 5c**, the ferroelectric layer adopts a P-Down polarization state. In this configuration, diode D1 becomes forward-biased while D2 remains reverse-biased. Although the ferroelectric polarization vector points toward the SRO/$PZTFN_{0.5}$ bottom interface, there can also exist a depolarization field ($E_{dp}$) with an opposing orientation, reducing the overall polarization. This polarizationinduced charge redistribution reduces the depletion width at the SRO/$PZTFN_{0.5}$ interface (black arrow) [25]. Finally, the overall electrical response is dominated by the forward-biased D1 diode, which explains the higher current observed in dark conditions for this polarization state (**Figure 4a**).

Under illumination ($h\nu \geq E_g$, 403 nm), photogenerated charge carriers separate according to the internal field: electrons migrate toward the top interface while holes drift toward the bottom interface. At zero external bias ($V_{ext} = 0$), the depolarization field oriented toward the top interface governs the electrical behavior, explaining the observed negative $J_{sc}$ values in **Figures 4b** and **4d**. Under dark conditions ($h\nu = 0$), the current remains at zero due to the absence of photogenerated charge carriers.

Under negative external bias ($V_{ext} < 0$, **Figure 5d**), the system exhibits an inverse configuration: the ferroelectric layer switches to the P-Up polarization state, and $E_{dp}$ points to the bottom interface, reversing the diode bias conditions (D1 becomes reverse-biased while D2 becomes forward-biased). In the P-Up state, the upward-oriented polarization vector reduces the depletion width at junction D1 below its forward-bias value (as indicated by the black arrow in the $PZTFN_{0.5}$/Au interface). The lower barrier height at junction D2 relative to D1 leads to decreased conductivity under these bias conditions. Under UV illumination, photogenerated electrons migrate to the bottom interface (D2) while holes are collected at the top interface (D1). At zero external bias ($V_{ext}$ = 0), the depolarization field again dominates the electrical behavior, resulting in the positive $J_{sc}$ shown in **Figures 4b** and **3d**. This comprehensive model successfully integrates the contributions of interface Schottky junctions and ferroelectric polarization dynamics, offering a consistent physical explanation for the observed switchable photovoltaic effect and current–voltage characteristics of the PZTFN$_{0.5}$ heterostructure.

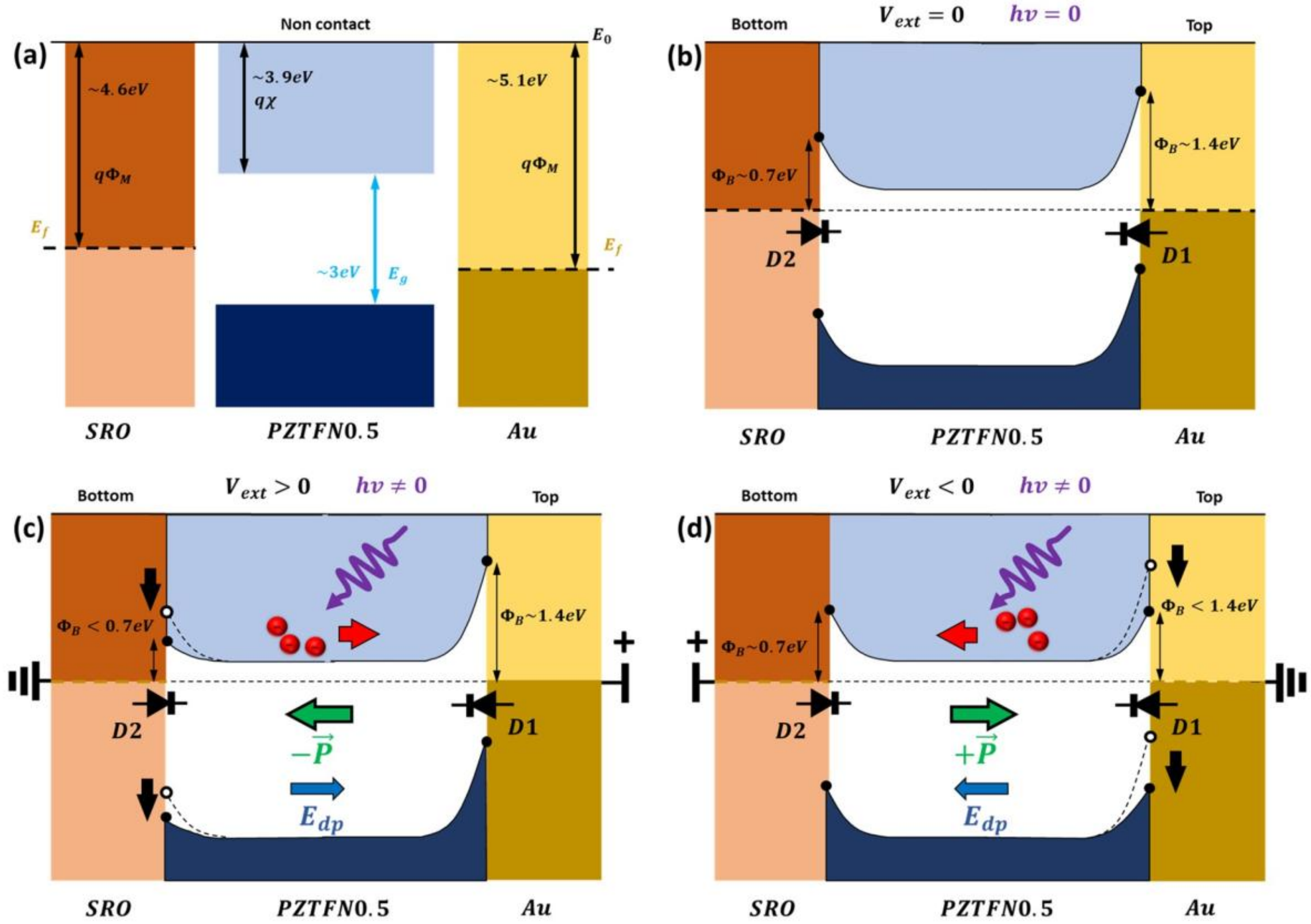

**Figure 5**. Schematic of the proposed band structure for the SRO/PZTFN$_{0.5}$/Au system. **(a)** Before contact, **(b)** just after contact, **(c)** contact condition with positive external field and light ON, and **(d)** contact condition with negative external field and light ON.

Finally, our findings demonstrate that Fe/Nb substitution in PZT significantly enhances the photoresponse of the films under 403 nm visible light illumination. This improved photocurrent, when compared with pure PZT films (see **Figure S6b**), directly correlates with an observed redshift of the optical band gap, which we attribute to the presence of Fe 3d midgap states [14]. It is also interesting to compare the photocurrent density values obtained for the $PZTFN_{0.5}$ film with those from $BiFeO_3$ (BFO), the most studied multiferroic material to date. For instance, unpoled, sputter-deposited BFO films (100 nm thick) with ITO electrodes yielded $J_{SC}$ of 4.5 $\mu$A/cm$^2$ under similar illumination conditions ($\lambda$ = 405 nm, 100 mW/cm$^2$) [26]. On the other hand, higher values, up to 25 $\mu$A/cm$^2$, have been reported for graphene/BFO/Pt heterojunctions [27], and optimized PLD BFO films have even reached ~ mA/cm$^2$ values [28]. Our $PZTFN_{0.5}$ films, although currently yielding smaller photocurrents, clearly highlight their potential to compete with BFO as a multiferroic photovoltaic material. In this context, our study provides a foundation for significantly enhancing the photoresponse of PZT-based thin films. Future work could focus on optimizing the Fe/Nb content or exploring additional A-site doping. For instance, the substitution of $Pb^{2+}$ by $Li^+$ and $Bi^{3+}$ at the A-site of $PZTFN_{0.5}$ appears to be a promising strategy to further increase the photovoltaic response [29].

## IV. CONCLUSIONS

We successfully grew high-quality PZT and 0.5PZT–0.5PFN ($PZTFN_{0.5}$) epitaxial films on $SrTiO_3$ (001) substrates by PLD, using $SrRuO_3$ as a conductive layer. The films exhibit a single perovskite phase with outstanding crystalline quality and nearly perfect (001) orientation. X-ray diffraction and reciprocal space mapping confirm their excellent structural alignment and minimal strain. However, TEM measurements reveal a confined region at the $PZTFN_{0.5}$/SRO interface, only a few nanometers thick, where dislocations and their associated strain fields are located.

The films exhibit well-defined, saturated hysteresis loops with a remanent polarization of 17 $\mu$C/cm$^2$ (PUND technique) and high coercive fields exceeding 100 kV/cm. Compared to polycrystalline films with the same composition and thickness, the epitaxial films display similar polarization and coercive field values. Magnetic measurements indicate weak ferromagnetic behavior at room temperature, with a remanent magnetization of 1.3 emu/cm$^3$ and a coercive field of 90 Oe.

The photovoltaic properties of Fe/Nb-substituted PZT films show promising performance under illumination, particularly when excited with a violet laser ($\lambda$ = 403 nm). The measured photocurrent density for the $PZTFN_{0.5}$ sample under violet light represents a significant improvement over PZT films, which exhibit no photocurrent response under similar conditions. Although the short-circuit photocurrent density ($J_{SC}$ = 23 $\mu A/cm^2$) and opencircuit voltage ($V_{OC}$ = −0.07 V) are smaller than those reported for BFO-based films, our results confirm a clear dependence between photocurrent and polarization, demonstrating the intrinsic coupling between ferroelectric domains and photovoltaic effects in the material. These findings highlight the potential of epitaxial (1−$x$)PZT–$x$PFN multiferroic thin films for photovoltaic applications, with significant opportunities for further enhancement of their photoresponse.

## CREDIT AUTHORSHIP CONTRIBUTION STATEMENT

Luc´ıa Imhoff: Data curation, Formal analysis, Investigation, Methodology, Validation, Writing – review editing. Miguel Andr´es Rengifo: Data curation, Formal analysis, Investigation, Methodology, Software, Supervision, Validation, Writing – review editing.Jos´e Manuel Caicedo Roque: Investigation, Resources. Jessica Padilla-Pantoja: Methodology, Resources. Jos´e Santiso: Funding acquisition, Investigation, Project administration, Resources, Supervision, Writing – review editing. Marcelo G. Stachiotti: Funding acquisition, Investigation, Project administration, Supervision, Writing – review editing. Myriam Hayd´ee Aguirre: Funding acquisition, Investigation, Project administration, Supervision, Validation, Writing – review editing.

## AI DECLARATION

During the preparation of this work, the authors used AI tools in order to enhance the manuscript's readability and writing quality. After using this tool, the authors reviewed and edited the content as needed and take full responsibility for the content of the publication.

**FUNDING**

This work was supported by the European Union's Horizon Europe Research and Innovation program under the Marie Sklodowska-Curie grant agreement N◦ 101110162 (NEMERFEC) and the Consejo Nacional de Investigaciones Científicas y Técnicas de la República Argentina (CONICET) through PIP N◦ 0374. M. G. S. Thanks support from the Consejo de Investigaciones de la Universidad Nacional de Rosario (CIUNR). We acknowledge the financial support of the European Commission through the H2020-MSCA RISE projects MELON (Grant N◦ 872631) and ULTIMATE-I (Grant N◦ 101007825).

**DATA AVAILABILITY STATEMENT**

The data that support the findings of this study are available from the corresponding author upon reasonable request.

**REFERENCES**

---

[1] T. Lottermoser, T. Lonkai, U. Amann, D. Hohlwein, J. Ihringer, and M. Fiebig, Nature (2004), 10.1038/nature02728.

[2] Z. Hu, G. B. Stenning, H. Zhang, Y. Shi, V. Koval, W. Hu, Z. Zhou, C. Jia, I. Abrahams, and H. Yan, Journal of Materiomics **11**, 100857 (2025).

[3] N. A. Spaldin and R. Ramesh, Nature Materials **18**, 203 (2019).

[4] T. Choi, S. Lee, Y. J. Choi, V. Kiryukhin, and S.-W. Cheong, Science **324**, 63 (2009).

[5] S. R. Basu, L. W. Martin, Y. H. Chu, M. Gajek, R. Ramesh, R. C. Rai, X. Xu, and J. L. Musfeldt, Applied Physics Letters **92** (2008), 10.1063/1.2887908.

[6] D. A. Sanchez, N. Ortega, A. Kumar, G. Sreenivasulu, R. S. Katiyar, J. F. Scott, D. M. Evans, M. Arredondo-Arechavala, A. Schilling, and J. M. Gregg, Journal of Applied Physics **113** (2013), 10.1063/1.4790317.

[7] D. Evans, A. Schilling, A. Kumar, D. Sanchez, N. Ortega, M. Arredondo, R. Katiyar, J. Gregg, and J. Scott, Nature Communications **4**, 1534 (2013).

[8] L. Imhoff, A. Rom´an, S. Barolin, N. Pellegri, L. Steren, M. Aguirre, and M. Stachiotti, Journal of the European Ceramic Society **42**, 2282 (2022).

[9] D. A. Sanchez, A. Kumar, N. Ortega, R. S. Katiyar, and J. F. Scott, Applied Physics Letters **97** (2010), 10.1063/1.3519979.

[10] D. A. Sanchez, K. K. Mishra, S. Saha, G. Srinivasan, and R. S. Katiyar, Crystals **13**, 1442 (2023).

[11] Y.-C. Wu, S. Z. Ho, Y. C. Liu, Y.-D. Liou, W.-Y. Liu, S.-W. Huang, J. Jiang, Y.-C. Chen, and J.-C. Yang, ACS Applied Electronic Materials **2**, 19 (2020).

[12] J. A. Schiemer, I. Lascu, R. J. Harrison, A. Kumar, R. S. Katiyar, D. A. Sanchez, N. Ortega, C. S. Mejia, W. Schnelle, H. Shinohara, A. J. F. Heap, R. Nagaratnam, S. E. Dutton, J. F. Scott, and M. A. Carpenter, Journal of Materials Science **51**, 10727 (2016).

[13] J. A. Schiemer, I. Lascu, R. J. Harrison, A. Kumar, R. S. Katiyar, D. A. Sanchez, N. Ortega, C. S. Mejia, W. Schnelle, H. Shinohara, A. J. F. Heap, R. Nagaratnam, S. E. Dutton, J. F. Scott, B. Nair, N. D. Mathur, and M. A. Carpenter, Journal of Materials Science **52**, 285 (2017).

[14] L. Imhoff, M. D. Marco, C. Lavado, S. Barolin, and M. Stachiotti, Journal of Alloys and Compounds **977**, 173428 (2024).

[15] Y. Zhou, H. K. Chan, C. H. Lam, and F. G. Shin, Journal of Applied Physics **98** (2005), 10.1063/1.1984075.

[16] V. Koval, G. Viola, and Y. Tan, "Biasing effects in ferroic materials," in *Ferroelectric Materials - Synthesis and Characterization* (InTech, 2015).

[17] M. Rath and M. S. R. Rao, Journal of Physics D: Applied Physics **52**, 244003 (2019).

[18] M. Fukunaga and Y. Noda, Journal of the Physical Society of Japan **77**, 1 (2008).

[19] J. F. Scott, L. Kammerdiner, M. Parris, S. Traynor, V. Ottenbacher, A. Shawabkeh, and W. F. Oliver, Journal of Applied Physics **64**, 787 (1988).

[20] F. Bern, M. Ziese, A. Setzer, E. Pippel, D. Hesse, and I. Vrejoiu, Journal of Physics: Condensed Matter **25**, 496003 (2013).

[21] J. C. Abrantes, J. A. Labrincha, and J. R. Frade, Materials Research Bulletin **35**, 727 (2000).

[22] G. Vats, Y. Bai, D. Zhang, J. Juuti, and J. Seidel, Advanced Optical Materials **7** (2019), 10.1002/adom.201800858.

[23] M. Zheng, X. Li, Q. Zhu, H. Li, L. Shi, X. Li, and R. Zheng, Superlattices and Microstructures **89**, 336 (2016).

[24] S. Sze and K. K. Ng, *Physics of Semiconductor Devices* (John Wiley & Sons, Inc., 2006).

[25] M. Rengifo, M. Aguirre, M. Sirena, U. Lu¨ders, and D. Rubi, Frontiers in Nanotechnology **4** (2022), 10.3389/fnano.2022.1092177.

[26] H. Chang, F. Yuan, Y. Yu, P. Chen, C. Wang, C. Tu, and S. Jen, Journal of Alloys and Compounds **574**, 402 (2013).

[27] Y. Zang, D. Xie, X. Wu, Y. Chen, Y. Lin, M. Li, H. Tian, X. Li, Z. Li, H. Zhu, T. Ren, and D. Plant, Applied Physics Letters **99** (2011), 10.1063/1.3644134.

[28] S. Sharma, M. Tomar, and V. Gupta, Vacuum **158**, 117 (2018).

[29] H. Borkar, V. Rao, M. Tomar, V. Gupta, J. Scott, and A. Kumar, Materials Today Communications **14**, 116 (2018).

# Enhanced Photocurrent Response in Epitaxial 0.5PZT–0.5PFN Multiferroic Thin Films

Luc´ıa Imhoff[1,2], Miguel A. Rengifo[3,4,5], Jos´e M. Caicedo Roque[6], Jessica Padilla-Pantoja[6], Jos´e Santiso[6], Marcelo G. Stachiotti[1], Myriam H. Aguirre[3,4,5]

[1]*IFIR - Instituto de F´ısica Rosario,*

*CONICET - Universidad Nacional de Rosario,*

*27 de Febrero 210 Bis, Rosario, Argentina*

[2]*FZU - Institute of Physics of the Czech Academy of Sciences, 182 00, Prague, Czechia*

[3] *Dept. de F´ısica de la Materia Condensada,*

*Universidad de Zaragoza, Pedro Cerbuna, 12, 50009, Zaragoza, Spain*

[4] *INMA - Instituto de Nanociencia y Materiales de Arag´on,*

*CSIC-Universidad de Zaragoza, Mariano Esquillor s/n, 50018, Zaragoza, Spain*

[5] *Laboratorio de Microscop´ıas Avanzadas, Universidad de Zaragoza,*

*Mariano Esquillor s/n, 50018, Zaragoza, Spain*

[6] *Catalan Institute of Nanoscience and Nanotechnology, ICN$^2$,*

*CSIC and The Barcelona Institute of Science and Technology (BIST),*

*Campus UAB, Bellaterra, Barcelona, Spain*

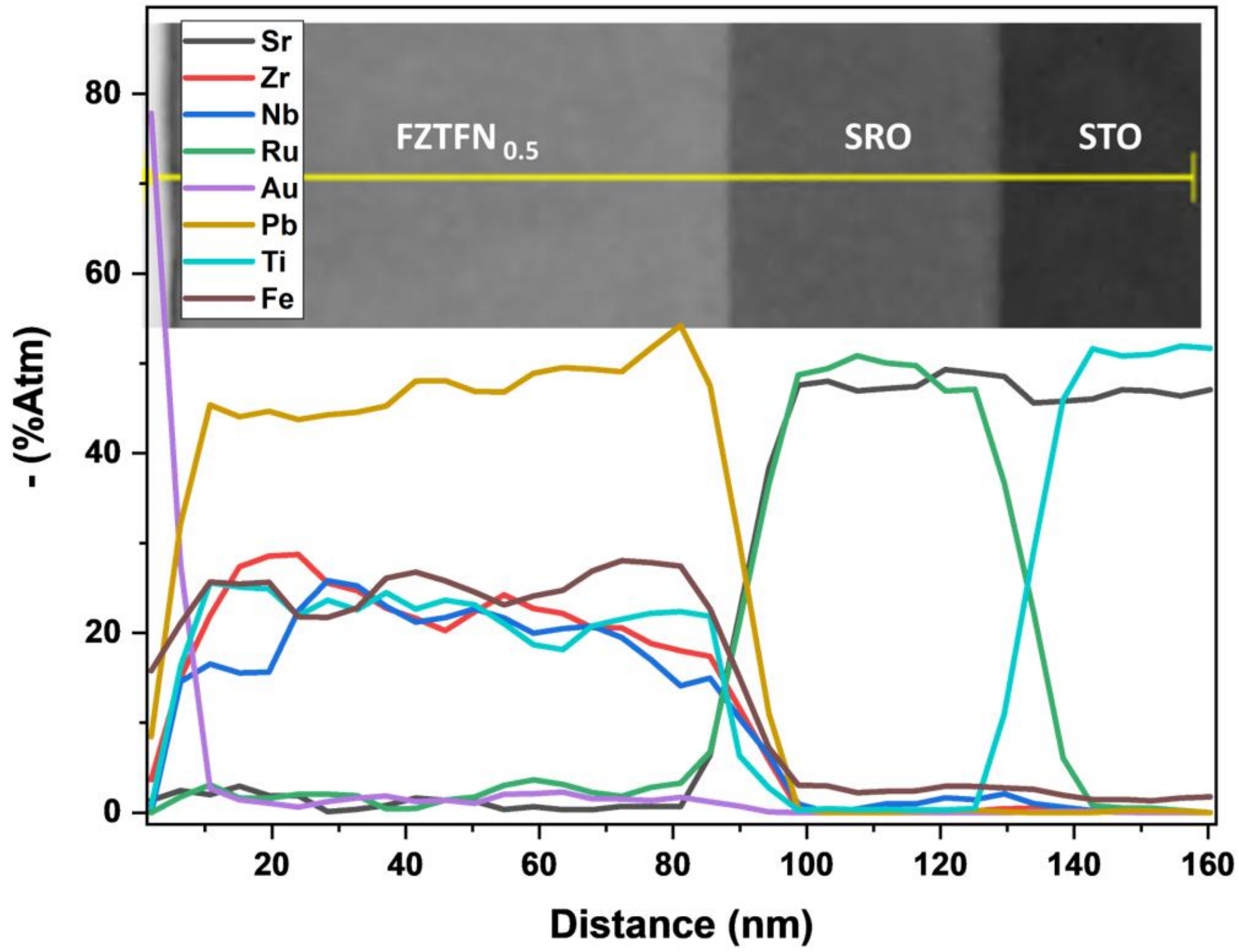

**Figure S1**. STEM-EDS line profile analysis across the epitaxial 0.5PZT–0.5PFN/SRO/STO heterostructure. The elemental distribution of Pb, Zr, Ti, Fe, and Nb confirms the presence of the multiferroic 0.5PZT–0.5PFN layer, while the Ru and Sr signals identify the SRO bottom electrode and the Ti/Sr signals correspond to the STO substrate. The inset shows the STEM-HAADF image indicating the region used for the compositional line scan.

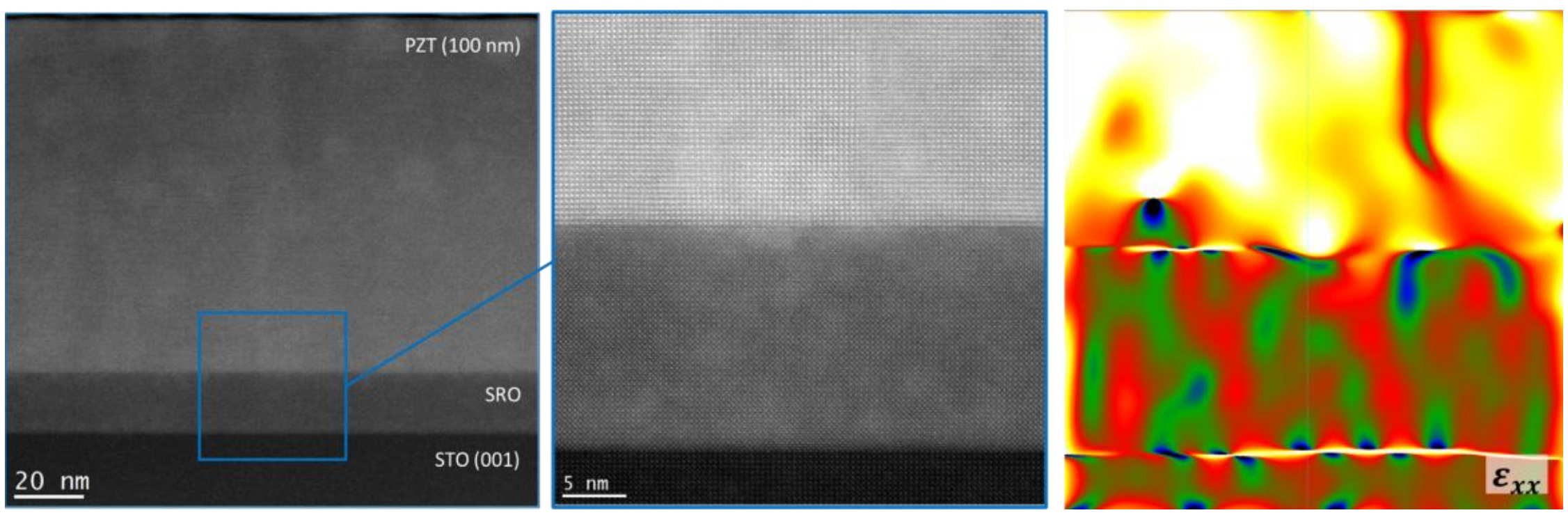

**Figure S2**. Low-magnification TEM image and atomic-resolution STEM-HAADF image of the STO/SRO/PZT heterostructure. The right panel shows the corresponding ($\varepsilon_{xx}$) strain mapping obtained from GPA analysis, evidencing the strain distribution near the SRO/PZT and SRO/STO interfaces.

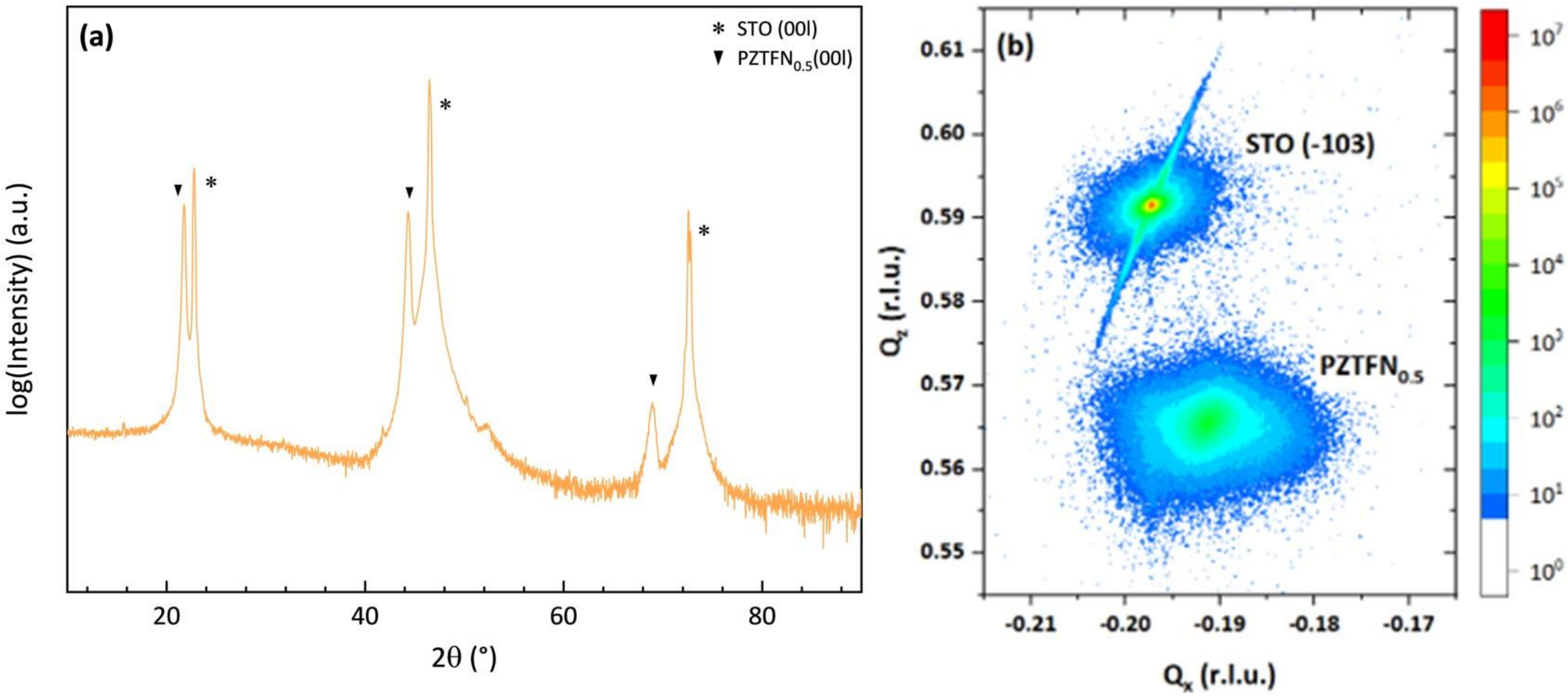

**Figure S3**. **(a)** Room-temperature XRD patterns of the STO//PZTFN$_{0.5}$ thin film. **(b)** Reciprocal space map (RSM) measured in the $(10\bar{3})$ reflection direction for the same sample. In contrast to the film grown on the SRO layer, when PZTFN$_{0.5}$ is deposited directly onto the STO substrate, it appears fully relaxed throughout the entire thickness. The fully relaxed state observed in the GPA strain maps is consistent with the X-ray RSM measurements presented previously in **Figure S2**.

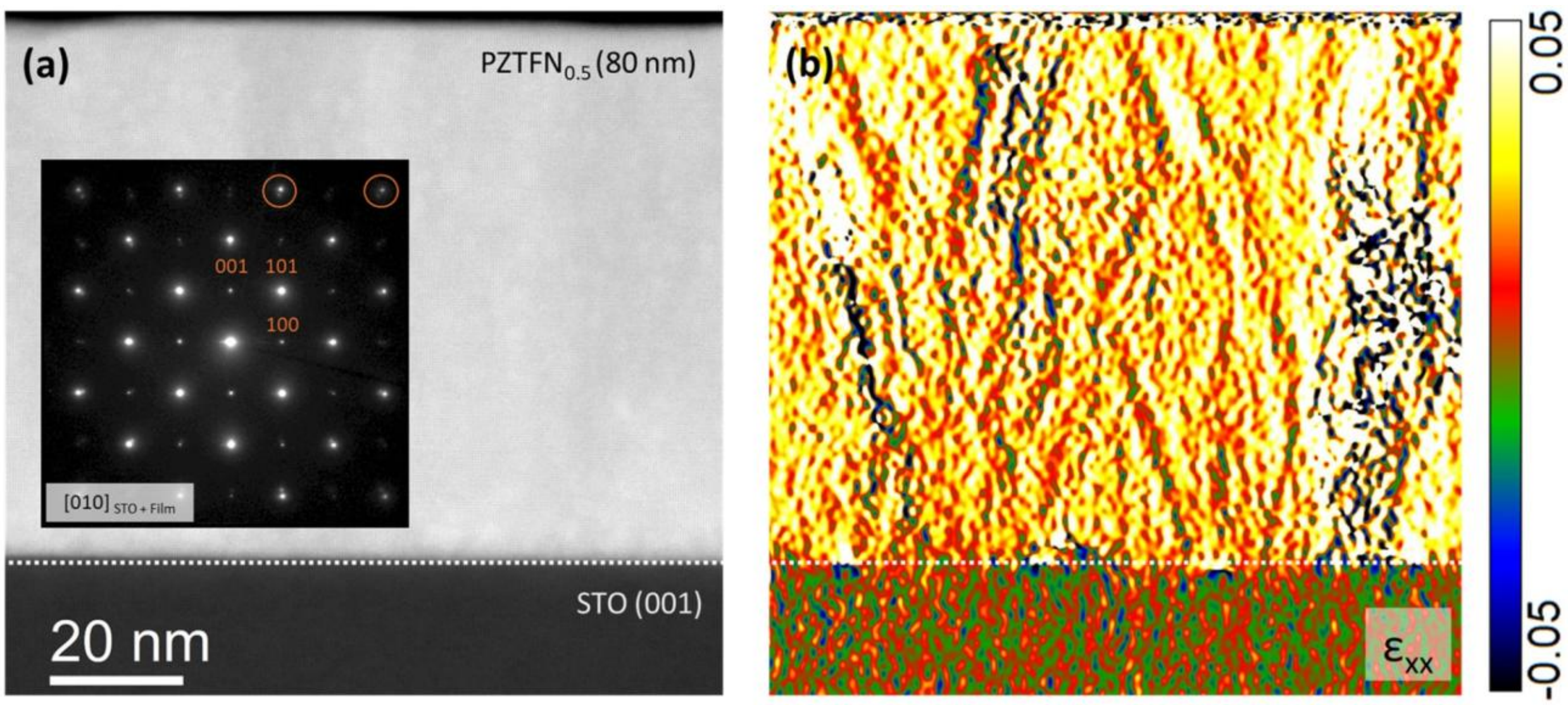

**Figure S4**. **(a)** Cross-sectional STEM-HAADF image of the STO//PZTFN$_{0.5}$ thin film, showing the STO substrate and a selected area electron diffraction (SAED) pattern taken from the interfacial region between the film and the STO substrate. Two sets of diffraction spots are visible, as indicated, due to the difference in lattice constants between the layers. **(b)** Cross-sectional image and strain ($\varepsilon_{xx}$) component map for the STO (001)//PZTFN$_{0.5}$ structure.

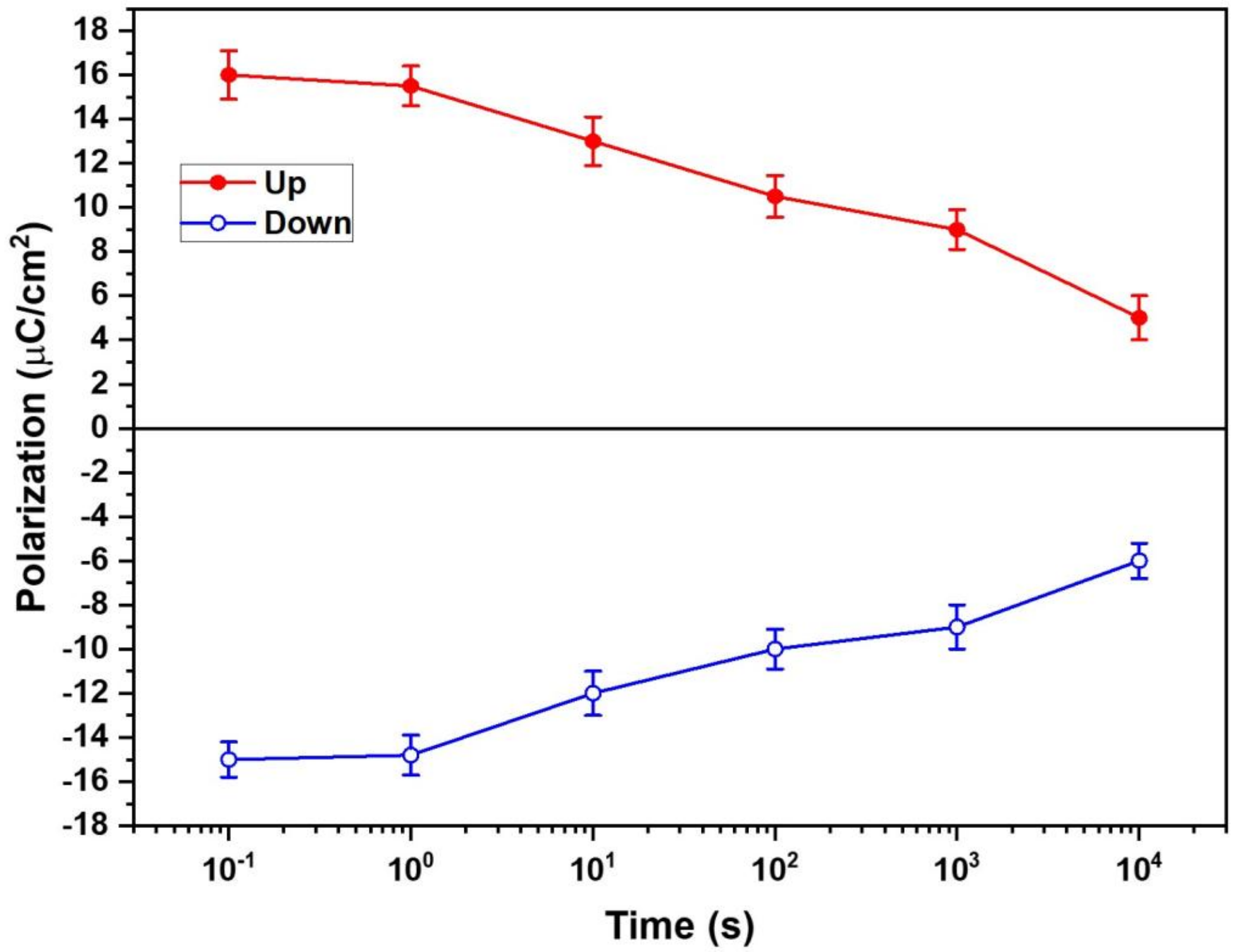

**Figure S5**. Polarization retention measurements obtained using the PUND method for the Up and Down polarization states of the epitaxial 0.5PZT–0.5PFN thin films. Both polarization states remain clearly distinguishable up to $10^4$ s, indicating stable ferroelectric retention behavior within the measured time window.

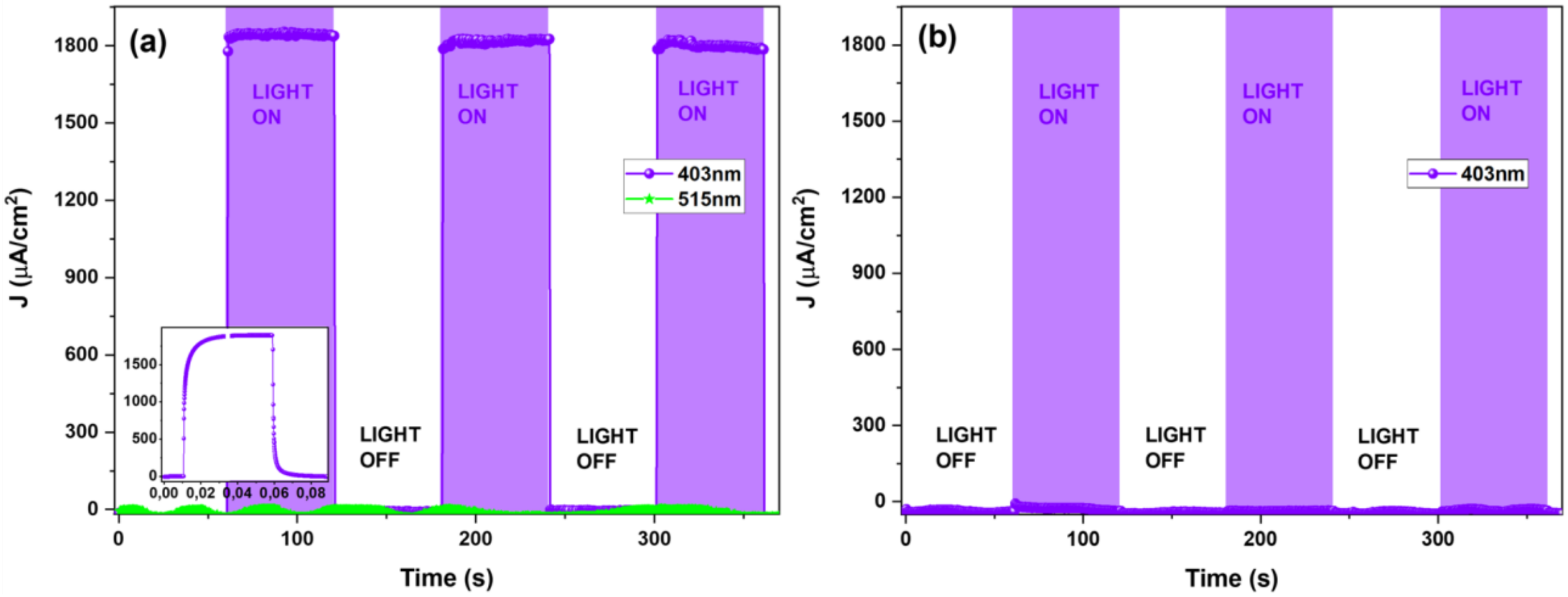

**Figure S6**. Photocurrent densities obtained under periodic light pulses for PZTFN0.5 **(a)** and reference PZT **(b)**. "Light on" corresponds to illumination, while "Light off" represents the measurement under dark conditions. Measurements were conducted at room temperature without any pre-poling treatment and with an applied voltage of 0.5 V. A clear increase in current density is observed when the PZTFN0.5 sample is illuminated with the violet light source of 403 nm. In contrast, no response is detected under green light illumination of 515 nm. For the reference PZT film, no significant photocurrent response is observed under 403 nm illumination. The inset in Figure **(a)** shows a higher temporal resolution measurement of the transient photocurrent response under 403 nm illumination, evidencing a fast excitation process followed by a slower relaxation behavior.